\documentstyle[aps,prl,twocolumn,epsf]{revtex}
\begin{document}
\draft

\twocolumn[\hsize\textwidth\columnwidth\hsize\csname @twocolumnfalse\endcsname

\title{The Chalker-Coddington Network Model is Quantum Critical}

\author{J. B. Marston and Shan-Wen Tsai}

\address{Department of Physics, Brown University, Providence, Rhode 
Island 02912-1843}

\date{\today}

\maketitle

\begin{abstract}
We show that the localization transition in the integer quantum Hall 
effect as described by the Chalker-Coddington network model is quantum 
critical.  We first map the anisotropic network model to the problem of  
diagonalizing a one-dimensional non-Hermitian non-compact 
supersymmetric lattice Hamiltonian of interacting bosons and fermions.  
Its behavior is investigated numerically using the density matrix 
renormalization group method, and critical behavior is found at the 
plateau transition.  This result is confirmed by a
generalization of the Lieb-Schultz-Mattis theorem.
\end{abstract}

\vskip1pc
]
Transitions between plateaus in the integer 
quantum Hall effect provide the clearest
example of quantum-critical behavior in a disordered system.  Understanding 
such critical points is a challenging problem because fluctuations occur over 
many decades of length and time scales, and averages over different 
realizations of the disorder must be carried out.  
Critical behavior was predicted by Levine, Libby, and Pruisken\cite{Pruisken}
and was observed experimentally
by Wei et al.\cite{Tsui} for temperatures close to the critical point
at absolute zero.  Progress toward a theoretical understanding of the plateau 
transition was achieved
with the introduction of a quantum tunneling network model 
by Chalker and Coddington\cite{Chalker}.  Subsequent numerical 
studies\cite{numerics} of the Chalker-Coddington model yielded values 
for the correlation length exponent $\nu \approx 2.3$ 
which were consistent with experiments.  
To the best of our knowledge, however, there has been no exact proof that the 
Chalker-Coddington model is quantum critical.

The method of supersymmetry (SUSY) can be used to analytically carry out 
disorder averages\cite{McKane,Efetov,Ian,Zirnbauer,Balents}.  We apply
the method to the anisotropic network model and then use the
Density Matrix Renormalization Group (DMRG) algorithm\cite{White} 
to study the resulting spin chain.  
Unlike usual spin chains such as the spin-$1/2$ Heisenberg 
antiferromagnet, the on-site Hilbert space in the SUSY chain is 
infinite-dimensional.  To access the 
critical point numerically therefore requires a double-extrapolation to 
large on-site Hilbert spaces and to large chain lengths.  
We use our numerical results to motivate
a generalization of the theorem of Lieb, Schultz and Mattis (LSM) which
confirms quantum criticality.

1. SUSY Spin-Chain for the Network Model.
The anisotropic Chalker-Coddington model can be 
represented\cite{Dung-Hai,Ziqiang} 
by an independent-particle Hamiltonian which describes a chain of $L$ (even)
edge states 
alternating in propagation forward and backward in imaginary time $\tau$. 
Random complex tunneling amplitudes $t_j(\tau)$ between adjacent edge states 
account for the Aharonov-Bohm phases accumulated by the electrons as they
circulate around equipotential contours of the random potential:
\begin{eqnarray}
H_{ip} = \int d\tau~ \left\{  \sum_{j=0}^{L-1}  
(-1)^j \psi^\dagger_j(\tau)~ {\rm i} \partial_\tau~ \psi_j(\tau) \right.
\nonumber \\
- \left. \sum_{j=0}^{L-2} [t_j(\tau) \psi^\dagger_j(\tau) \psi_{j+1}(\tau) 
+ t^*_j(\tau) \psi^\dagger_{j+1}(\tau) \psi_j(\tau)] \right\} .
\label{anisotropic}
\end{eqnarray}
The $(-1)^j$ term has its origin in the alternating
propagation of adjacent edge states and this factor reappears several times
in the equations which follow.
The disorder average of the tunneling amplitudes is given by
\begin{eqnarray}
\overline{t_j^*(\tau)~ t_{j^\prime}(\tau^\prime)} 
&=& J_{j}~ \delta_{j,j^\prime}~ \delta(\tau - \tau^\prime)~ ,
\nonumber \\
J_j &=& [1+(-1)^j R]\ .
\label{variance}
\end{eqnarray}
The relevant dimerization parameter $R = \pm 1$ deep inside the plateaus; 
the transition between the plateaus occurs at $R = 0$.   
Disorder averaging of the corresponding functional integral is made possible 
with the use of SUSY\cite{Zirnbauer2,Kondev} 
as the partition function $Z = 1$ for
each realization of the disorder.  Transfer matrix formalism can be used to 
resolve normal-ordering ambiguities\cite{Ilya,Mathur} and the resulting 
effective SUSY Hamiltonian may then be extracted.  It describes  
interacting spin-up and spin-down fermions $c_\sigma$ and bosons 
$b_\sigma$ (two spin species
are introduced to permit the calculation of the disorder-averaged 
product of retarded and advanced Green's functions which determines the 
conductivity): 
\begin{eqnarray}
H &=& \sum_{j=0}^{L-2}~ J_j~ 
\left[ \sum_{a=1}^8~ g_a~ S^a_j~ S^a_{j+1}
 + (-1)^j~ \sum_{a=9}^{16}~ g_a~ S^a_j~ S^a_{j+1} \right]   
\nonumber \\
&& + \eta \sum_{j=0}^{L-1} \left[ S^1_j + S^2_j + S^5_j + S^6_j \right]\ . 
\label{H-SUSY}
\end{eqnarray}
Parameter $\eta > 0$ ensures convergence of the non-compact bosonic sector and
defines the advanced and retarded propagators; also the signs 
$g_a$ are given by:
\begin{eqnarray}
g_a = \left\{ \begin{array}{ll} 1;~ a = 1, 2, 10, 12, 14, 16\\
\\
-1;~ a = 3, \ldots, 9, 11, 13, 15\ .
\end{array}\right.
\label{signs}
\end{eqnarray}
In Eq. \ref{H-SUSY} we have introduced 16 spin operators, the components of
a $4 \times 4$ superspin matrix:
\begin{eqnarray}
\begin{array}{l} 
S^1 \equiv b^\dagger_\uparrow b_\uparrow + 1/2 \\ \\
S^2 \equiv b^\dagger_\downarrow b_\downarrow + 1/2 \\ \\
S^3 \equiv b^\dagger_\uparrow b^\dagger_\downarrow \\ \\
S^4 \equiv b_\downarrow b_\uparrow \\ \\
S^{5} \equiv c^\dagger_\uparrow c_\uparrow - 1/2 \\ \\
S^{6} \equiv c^\dagger_\downarrow c_\downarrow - 1/2 \\ \\
S^{7} \equiv c^\dagger_\uparrow c^\dagger_\downarrow \\ \\
S^{8} \equiv c_\downarrow c_\uparrow \\ \\
\end{array} \ \ \  
\begin{array}{l}  
S^9 \equiv c^\dagger_\downarrow b_\downarrow \\ \\ 
S^{10} \equiv c^\dagger_\uparrow b_\uparrow \\ \\
S^{11} \equiv b^\dagger_\downarrow c_\downarrow \\ \\
S^{12} \equiv b^\dagger_\uparrow c_\uparrow \\ \\ 
S^{13} \equiv b_\downarrow c_\uparrow \\ \\ 
S^{14} \equiv b_\uparrow c_\downarrow \\ \\
S^{15} \equiv b^\dagger_\downarrow c^\dagger_\uparrow \\ \\
S^{16} \equiv b^\dagger_\uparrow c^\dagger_\downarrow\ . \\ \\
\end{array} 
\label{superspins}
\end{eqnarray}
Bosonic-valued operators $S^1, \ldots, S^8$ make up the symmetric sector of the
Hamiltonian while fermion-valued operators $S^9, \ldots, S^{16}$ are in the 
antisymmetric sector.  Despite the fact that $H$ is non-Hermitian, it 
only has real-valued eigenvalues.  

The Hamiltonian commutes with four (fermion-valued) supersymmetry generators, 
$[H,~ Q_{1 \sigma}] = [H,~ Q_{2 \sigma}] = 0$, where  
\begin{eqnarray}
Q_{1 \sigma} &\equiv& \sum_j \bigg{[} b^\dagger_{j \sigma} c_{j \sigma}
- (-1)^j c^\dagger_{j \sigma} b_{j \sigma} \bigg{]} .
\nonumber \\
Q_{2 \sigma} &\equiv& \sum_j \bigg{[} (-1)^j b^\dagger_{j \sigma} 
c_{j \sigma} + c^\dagger_{j \sigma} b_{j \sigma} \bigg{]} .
\label{SUSY-charges}
\end{eqnarray}
It is not difficult to see that the supersymmetric Hamiltonian must have a 
unique, zero-energy, ground state.  The right and left (ground) eigenstates 
are therefore annihilated by the Hamiltonian:  
$H | \Psi_0 \rangle = \langle \Psi_0 | H = 0$.  Also, the
ground state is annihilated by the SUSY charges:
$Q_{1 \sigma} | \Psi_0 \rangle = Q_{2 \sigma} | \Psi_0 \rangle = 0$. 
All excited states appear in quartets or 
larger multiples of 4, half with odd total fermion content, 
and these cancel out in the partition function by virtue 
of the supertrace:
\begin{equation}
Z = {\rm STr} e^{-\beta H} \equiv {\rm Tr} (-1)^{N_c} e^{-\beta H} = 1, 
\label{partition}
\end{equation}
where $N_c$ is the total number of fermions.
In the limit of $\eta \rightarrow \infty$, 
$|\Psi_0 \rangle \rightarrow | 0 \rangle$.  For finite $\eta > 0$, 
however, it is a remarkable fact that the normalized
ground state is a superposition of the vacuum state
with {\it unit} amplitude and an infinite number of {\it zero-norm}
many-body states $| J \rangle$ with differing total number of 
particles\cite{Balents2}: 
\begin{equation}
|\Psi_0 \rangle = | 0 \rangle + \sum_{J=1}^\infty a_J(\eta)~ 
| J \rangle;\ \ \ \ \langle I | J \rangle = 0\ \forall\ I, J > 0.
\label{ground-content}
\end{equation}
This result can be verified directly by observing that 
$| \Psi_0 \rangle = \lim_{\beta \rightarrow \infty} e^{- \beta H} | 0 \rangle$
and application of powers of $H$ to the vacuum state yields only  
zero-norm states.  We have also checked numerically, for finite length chains, 
that the vacuum state has unit amplitude 
when the Hilbert space is truncated in a way which respect supersymmetry 
(see below).
It is useful to contrast the complicated ground state of the non-Hermitian
SUSY Hamiltonian with the ground state of the {\it Hermitian}
SUSY ferromagnet which describes a chiral metal
with all edge states propagating in the same direction\cite{Balents}. 
Backscattering is absent,
localization cannot occur, and the ground state of the SUSY ferromagnet is 
simply the vacuum state.   

2. DMRG Analysis.  
To simply show that the density of states (DOS) 
is non-vanishing it suffices to remove
one of the spin sectors, for example the $\downarrow$-spins.  
The remaining $\uparrow$-spin 
degrees of freedom are then compact, the ground state is 
the vacuum state $|0 \rangle$, the DOS is proportional to 
$\langle S^1 \rangle = - \langle S^5 \rangle = 1/2$, 
and the Hamiltonian can be exactly diagonalized\cite{Kondev}.   
When both spin sectors are included the theory 
is non-compact and highly non-trivial.  To make further progress we employ the 
infinite-size DMRG method\cite{White}.  
The Hilbert space is first constructed systematically on each site 
by repeated action of the double-creation operator 
$S^3 \equiv b^\dagger_\uparrow b^\dagger_\downarrow$. 
Introducing the integer level index $n = 0, 1, 2, \ldots$ we add to
the vacuum state $|0 \rangle$ a tower of states built out of the quartets: 
\begin{eqnarray}
|4 n + 1 \rangle &\equiv& \frac{1}{n!}
(b^\dagger_\uparrow b^\dagger_\downarrow)^n 
c^\dagger_\uparrow c^\dagger_\downarrow |0 \rangle 
\nonumber \\ 
|4 n + 2 \rangle &\equiv& \frac{1}{\sqrt{n! (n+1)!}}
(b^\dagger_\uparrow b^\dagger_\downarrow)^n 
b^\dagger_\uparrow c^\dagger_\downarrow |0 \rangle 
\nonumber \\ 
|4 n + 3 \rangle &\equiv& \frac{1}{\sqrt{n! (n+1)!}}
(b^\dagger_\uparrow b^\dagger_\downarrow)^n
c^\dagger_\uparrow b^\dagger_\downarrow |0 \rangle 
\nonumber \\ 
|4 n + 4 \rangle &\equiv& \frac{1}{(n+1)!}
(b^\dagger_\uparrow b^\dagger_\downarrow)^n 
b^\dagger_\uparrow b^\dagger_\downarrow |0 \rangle \ .
\label{Hilbert}
\end{eqnarray}
Truncations with $D = 4 n + 1$ states preserve supersymmetry, as the SUSY
generators, Eq. \ref{SUSY-charges}, intermix 
the quartet of states, Eq. \ref{Hilbert}, separately within 
each level of the tower
without changing the total number of particles. The DOS remains unchanged, 
and the ground 
state energy is exactly zero, providing a valuable check on the accuracy of 
the DMRG algorithm which incurs errors when, as the chain length increases, the 
Hilbert spaces of the blocks grow beyond the finite limit of $M$ states.
Increasing $M$ up to limits set by machine memory and speed 
yields systematic improvement in the accuracy of the DMRG 
algorithm.  In results reported below we have checked that $M$ is sufficiently
large to ensure adequate accuracy; even in the challenging case of 
$D=13$ and $M=170$ the ground state, when targeted, had an energy which 
deviated from zero by less than $0.003$. 

\begin{figure}
\epsfxsize=3.5in \epsfbox{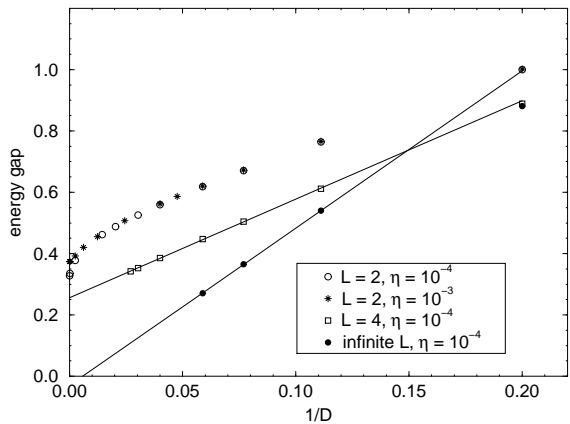}
\caption{Gaps to the lowest excited state for chains with open boundary
conditions and $R = 0$.  The gap for the two-site
problem was obtained by diagonalizing Eq. 61 of Ref. 16. A straight line is 
fit to the $L = 4$ points (with $M = D$).  
Also shown are gaps at $L \rightarrow \infty$ 
which are obtained from the extrapolations presented in Fig. 2 
for the supersymmetric truncations $D = 5, 9, 13$, and $17$.
A straight line is fit to the last three points.} 
\label{extrapolations}
\end{figure}

\begin{figure}
\epsfxsize=3.5in \epsfbox{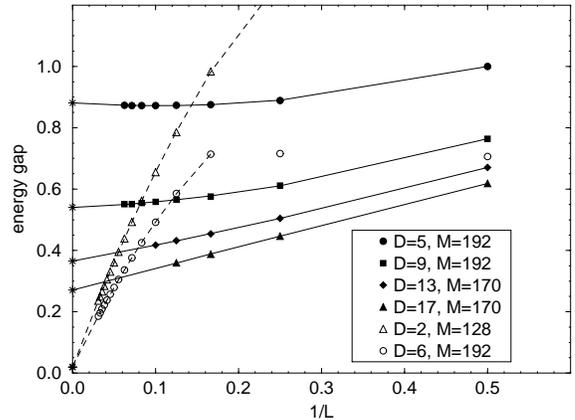}
\caption{Gaps, for $\eta = 10^{-4}$, extrapolated to 
$L \rightarrow \infty$.  Truncations which respect supersymmetry 
(solid lines) and non-supersymmetric truncations (dashed lines) are plotted.  
Points at $L = \infty$ are obtained by fitting the gap to the form 
$a + b/L + c/L^2$.}
\label{length}
\end{figure}

Reduced density matrices for the two augmented blocks,
each of Hilbert space size $D \times M$, are formed by 
computing a partial trace over half the chain.  
For the left half of the chain the density matrix is given by: 
\begin{equation}
\rho_{i j} = \sum_{i^\prime=1}^{D M}~ \Psi_{i i^\prime}~ \Psi_{j i^\prime}.
\label{density}
\end{equation}
Here $\Psi_{i i^\prime} \equiv \langle i, i^\prime | \Psi \rangle$
are the real-valued matrix elements of the targeted 
many-body wavefunction projected onto a basis
of states labeled by unprimed Roman index $i$ which covers the left half
of the chain and primed index $i^\prime$ which covers the right half.
To compute the gap, $\Psi$ is chosen to
be one of the lowest-lying right eigenstates of $H$.
All of the eigenvalues of $\rho$ are real and positive;
these are interpreted as probabilities and the $(D-1) M$ least probable 
states are thrown away.  The desired $D \rightarrow \infty$ limit can
be reached by extrapolation, Fig. \ref{extrapolations}.   
Fig. \ref{length} shows that the gap is non-zero in the
thermodynamic limit for finite fixed $D = 4 n + 1$ (solid lines), but as 
shown in Fig. \ref{extrapolations} it approaches zero
in the $D \rightarrow \infty$ limit as expected.

Also of interest are non-supersymmetric truncations $D = 4 n + 2$, with 
the state $| 4 n + 1 \rangle$ selected as the final state at the top of 
the tower.  The special
case of no bosons, $D=2$, with on-site states $\{ | 0 \rangle,~ 
c^\dagger_\uparrow c^\dagger_\downarrow | 0 \rangle \}$ 
corresponds to the ordinary spin-$1/2$ Heisenberg 
antiferromagnet as can be verified by making separate particle-hole 
transformations on the even and odd sublattices:  
$c_{2 j \uparrow} \leftrightarrow c^\dagger_{2 j \uparrow}$ and
$c_{2 j + 1 \downarrow} \leftrightarrow c^\dagger_{2 j + 1 \downarrow}$;
consequently the gap vanishes in this case.  As seen in 
Fig. \ref{length} the gap also vanishes for $D = 6$.  We 
prove below that all truncations with $D = 4n + 2$ are gapless.

3. Lieb-Schultz-Mattis Theorem. 
For half-odd-integer spin antiferromagnets on a periodic chain of 
length $L$ sites ($\vec{S}_L \equiv \vec{S}_0$) 
LSM showed\cite{LSM} that either 
(1) the ground state is degenerate or (2) there are 
gapless spin excitations in the thermodynamic limit $L \rightarrow \infty$.  
LSM employed 
a variational argument by introducing the unitary slow-twist operator, 
\begin{equation}
U \equiv \exp \left\{ \frac{ 2 \pi i}{L}~ \sum_{j=0}^L j S^z_j \right\} , 
\label{LSMU}
\end{equation} 
which has the property that $U^\dagger [H, U] = O(1/L)$.  Now 
$\langle \Psi_0 | U |\Psi_0 \rangle = 0$ because 
$U \rightarrow -U$ under 
parity of reflection about the middle site ($j \leftrightarrow L - j$) 
combined with a rotation of 180 degrees 
about the y-axis ($S^z \leftrightarrow -S^z$). 
So $U$ either creates a low-energy 
excitation above a unique ground state or mixes degenerate ground states.  
In contrast, for integer spins $U \rightarrow U$ under parity and $U$ does not,
in general, create a low-energy excitation or switch degenerate ground states. 

In the SUSY problem we are able to make a stronger statement, because we
know that the ground state is unique by supersymmetry.
The natural generalization of the LSM slow-twist operator for the SUSY chain is:
\begin{equation}
U \equiv \exp \left\{ \frac{\pi i}{L}~ \sum_{j=0}^L j (-1)^j 
[ n_c(j) + n_b(j) - 1] \right\}, 
\end{equation} 
where $n_b(j) \equiv b^\dagger_{j \sigma} b_{j \sigma}$ is the number of 
bosons on site $j$ and 
$n_c(j) \equiv c^\dagger_{j \sigma} c_{j \sigma}$ is the number of fermions. 
It reduces, in the $D = 2$ limit of no bosons, and after the particle-hole
transformation is taken, to the usual LSM operator Eq. \ref{LSMU}.  
In the ground state, the sum $n_c(j) + n_b(j)$
is always an even number, so $U$ respects the periodic boundary condition and
it also has the desired property $U^\dagger [H, U] = O(1/L)$.  
The canonical parity transformation now takes the form:
\begin{eqnarray}
j &\leftrightarrow& L - j
\nonumber \\
(-1)^j &\leftrightarrow& (-1)^j
\nonumber \\
c^\dagger_{j \sigma} &\leftrightarrow& c_{L-j \sigma}
\nonumber \\ 
b_{j \uparrow} &\leftrightarrow& b_{L-j \downarrow}
\label{susy-parity}
\end{eqnarray}
again reducing in the absence of bosons to the usual LSM parity operation.  The
supersymmetric Hamiltonian is invariant under this operation, 
but only at the presumed critical point $R = \eta = 0$.  The slow-twist
operator $U$, however, changes form because while $n_c \rightarrow 
2 - n_c$, the number of bosons $n_b$ remains invariant:
\begin{equation}
U \rightarrow -\exp \left\{ \frac{\pi i}{L}~ \sum_{j=0}^L j (-1)^j 
[ n_c(j) - n_b(j) - 1] \right\}. 
\label{change-form}
\end{equation} 
However, $U$ is invariant under global supersymmetry rotations, 
$[Q_{1 \sigma},~ U] = [Q_{2 \sigma},~ U] = 0$.  As only the ground state is a 
SUSY singlet, $U$ cannot create a
low-energy excitation for truncations which respect SUSY,  
consistent with the above DMRG results.  Indeed, from Eq. \ref{ground-content} 
it follows that $| \langle \Psi_0 | U | \Psi_0 \rangle | = 1$ and thus 
$U | \Psi_0 \rangle$ does not contain a component orthogonal to  
the ground state $| \Psi_0 \rangle$.  (This can be viewed as an alternative 
proof of Eq. \ref{ground-content}.)  For non-SUSY
truncations $D = 4 n + 2$, however, the ground state does not obey 
Eq. \ref{ground-content}; instead $|\langle \Psi_0 | U | \Psi_0 \rangle| < 1$
as can be verified either directly for small chains, and in the special 
case $D = 2$ of no bosons 
(for which $\langle \Psi_0 | U | \Psi_0 \rangle = 0$), or by appealing to the
fact that $U$ changes form under the parity operation, 
Eq. \ref{change-form}.  For sufficiently large $D$ and $\eta > 0$ the 
ground state approaches the SUSY ground state and is thus unique; 
therefore $U$ creates low-energy excitations\cite{non-negative}, and 
chains with the non-SUSY truncation $D = 4 n + 2$ are gapless
in the thermodynamic limit. 

Before examining the physically relevant 
$D \rightarrow \infty$ limit, first consider the 
large spin limit of one-dimensional nearest-neighbor Heisenberg 
antiferromagnets.  Chains with spin $S = n + 1/2$ (and even-numbered 
$D = 2 S + 1 = 2 n + 2$) are gapless for all integer $n \geq 0$.  
Chains with $S = n + 1$ (and odd-numbered $D = 2 n + 3$) are 
Haldane gapped, but this gap must
vanish in the limit of large spin to accord with the gaplessness of
the half-odd-integer in the $n \rightarrow \infty$ limit.  This reasoning
can be checked by a simple renormalization-group argument using the
beta function for the  
non-linear $\sigma$-model which shows that the gap for integer-spin chains
vanishes as $e^{-\pi S}$. Likewise, for $\eta = 0^+$, continuity requires 
that SUSY truncations with odd-numbered $D = 4 n + 1$ must converge to the 
gapless behavior exhibited by the non-SUSY even-numbered $D = 4 n + 2$ 
truncations in the $n \rightarrow \infty$ limit.  Hence the 
Chalker-Coddington model is quantum critical at $R = 0$. 

We thank W. Craig, S. Das Sarma, M.P.A. Fisher, I. Gruzberg, G. Guralnik,
V. Gurarie, B. Huckestein, J. Kondev, A. W. W. Ludwig, T. Senthil, and 
Z. Wang for helpful comments.   
J.B.M. thanks the Institute for Theoretical Physics for its hospitality
during the completion of this work. 
This research was supported in part by the NSF under
Grants Nos. DMR-9357613, DMR-9712391 and PHY94-07194.  Computations were
carried out in C++ on Cray PVP machines at 
the Theoretical Physics Computing Facility at Brown University.


\begin{references}
\bibitem{non-negative} The state $U | \Psi_0 \rangle$ is a superposition of
eigenstates, each of non-negative norm, because the non-SUSY truncated 
Hamiltonian may be viewed as the SUSY truncation plus a Hermitian 
perturbation.  The zero-norm SUSY states acquire non-negative norm and 
an upper bound on the energy of the excitation can be established.
\end{references}
\end{document}